\documentclass[12pt]{article}
\newcommand{\be}{\begin{equation}}
\newcommand{\ee}{\end{equation}}
\newcommand{\ba}{\begin{eqnarray}}
\newcommand{\ea}{\end{eqnarray}}
\newcommand{\ci}[1]{\cite{#1}}

\newcommand{\la}[1]{\label{#1}}

\def\gl#1{(\ref{#1})}

\def\Tr#1{\mbox{\rm Tr}\left(#1\right)}

\renewcommand{\thefootnote}{\fnsymbol{footnote}}
\date{}

\begin{document}

\begin{center}

{\Large\bf Supersymmetrization of the Franke-Gorini-Kossakowski-Lindblad-Sudarshan equation.}

\vspace{1cm}
{\large A. A. Andrianov\footnote{E-mail: a.andrianov@spbu.ru}, M. V. Iof\/fe\footnote{E-mail: m.ioffe@spbu.ru}, O. O. Novikov\footnote{E-mail: o.novikov@spbu.ru}}
\\
\vspace{0.5cm}
Saint Petersburg State University, 7/9 Universitetskaya nab., St.Petersburg, 199034 Russia.\\

\end{center}

\vspace{1cm}

\begin{abstract}
We investigate departures from the Hamiltonian dynamics such as the quantum decoherence in superpartner quantum systems due to their interaction with the environment.
The effective description is given by the Franke-Gorini-Kossakowski-Lindblad-Sudarshan (FGKLS) equation.
The prescription for supersymmetrization of this equation is elaborated and  pairs of related  isospectral quantum
systems are built. We present some instructive examples of such a construction.
\end{abstract}
\bigskip
\renewcommand{\thefootnote}{\arabic{footnote}}

\section{Introduction}

The interaction between the quantum system of interest with its environment causes the departures from the Hamiltonian dynamics.
The resulting evolution is typically described by the Franke-Gorini-Kossakowski-Lindblad-Sudarshan (FGKLS) master equation \cite{franke} -
\cite{lindblad-2} for the density matrix $\rho .$
This equation takes into account the effects of the environment such as the decoherence using the extra non-Hamiltonian term constructed from the so called FGKLS operators while preserving the linearity and the probability conservation.

One remark on the title of the present paper is in due here. Recently, in their paper \cite{history} "A Brief History of the GKLS Equation", D.Chruscinski and S.Pascazio reconstructed chronologically "the chain of events, intuitions and ideas that led" V.Gorini, A.Kossakowski, G.Lindblad and G.Sudarshan to the formulation of so-called GKLS equation in \cite{sudar} and \cite{lindblad}. We have to inform the readers that one more publication should be included in the row. We mean the paper \cite{franke} of V.A.Franke entitled "On the general form of the dynamical transformation of density matrices". There was studied the most general linear transformations of the density matrices of a quantum system that preserve the total probability and do not generate negative probabilities in this paper. As a result, the master equation (a little bit more general than "Lindblad equation" of \cite{sudar}, \cite{lindblad}) for evolution of the density matrix was derived. The paper of V.A.Franke was submitted to journal on October, 1975, it was published in Russian on May, 1976, and on the same 1976, it appeared in English version \cite{franke} {\it Theoretical and Mathematical Physics} \footnote{Subsequently, V.A.Franke worked on the relativistic generalization of the Lindblad equation \cite{franke-2}.}.

We assume the sufficiently weak coupling of the system to the environment with large number of degrees of freedom resulting in the irreversible Markovian evolution.
There are many examples of such systems in the atomic and
solid state physics and optics\cite{many} - \cite{many-3} (an updated list of applications see in \cite{albert} - \cite{albert-3}).
An important application in the particle physics is provided by the neutral
kaon-antikaon propagation which can be influenced by the vacuum
quantum-gravity fluctuations \cite{PH} - \cite{E-2} and/or by the detector matter
\ci{ATT}. In his recent papers, Steven Weinberg gave new impetus to the discussion of
the $\rho$ matrix approach as the only ingredient necessary to incorporate the measurement and the FGKLS equation to be a smooth way to describe the relevant loss of quantum coherence after the influence of device \cite{weinberg}, \cite{weinberg-2}.

 Correspondingly the consistent description is based on the formalism of the $\rho$ matrix
rather than on the wave functions themselves as the decoherence converts pure
states with a given wave function
into mixed ones which are characterized not by a particular
wave function $\psi$ but by a density matrix,
\begin{equation}
\rho=\rho^\dagger,\quad \langle\phi|\rho|\phi\rangle\geq 0,
\quad \forall |\phi\rangle.
\end{equation}
The expectation values of the observables $A$ are obtained using the relation,
\be
\langle A\rangle \equiv \Tr{\hat{A}\rho}.
\ee
This implies that $\Tr{\rho}$ plays the role of the total probability. If it is conserved it is assumed to be normalized to,
\be
\Tr{\rho}=1.\la{2}
\ee

In the standard (Hamiltonian) Quantum Mechanics, Supersymmetrical approach \cite{reviews} - \cite{reviews-5} gave a powerful impulse in construction of exactly solvable and quasi exactly
solvable quantum models of different nature \cite{nature} - \cite{nature-10}. The main elements of Supersymmetrical Quantum Mechanics (SUSY QM) are the SUSY intertwining relations which express the commutation of Superhamiltonian with Supercharges. They provide the connection between pairs of superpartner systems: spectral properties of these systems are related to each other. It is possible to iterate this procedure to construct the chain of partner systems connected by the supertransformations of different orders \cite{reviews-4}, \cite{ais} -
\cite{levels-2}. Thus, the SUSY method in Quantum Mechanics became one of the best tools for quantum design \cite{levels}, \cite{levels-2}. It looks interesting to generalize the SUSY approach for quantum systems with Hamiltonian dynamics to a more general case of quantum systems interacting with environment and described by the FGKLS  equation. Just this problem is studied in the present paper. It is organized as follows. After the brief discussion of the FGKLS equation in Section 2, the prescription for supersymmetrization of this equation is proposed in Section 3. Finally, Section 4 contains few illustrating examples of such construction.

\section{FGKLS equation}
The evolution of the {\it total} system [Subsystem + (Large) Environment]
is unitary, with the time evolution operator
given by $U(t)=\exp(-i H_{Total}t)$;
whereas the dynamics of the sub-system alone is obtained by
tracing out the environment degrees of freedom:
\begin{equation}
\rho_{SUB}(t)= {\rm Tr}_{ENV} \left( U(t) \rho_{SUB+ENV}(0)
U^\dagger (t) \right).
\label{evolution}
\end{equation}
This gives a complicated evolution for $\rho_{SUB}$,
even if the initial conditions
are a direct product $\rho_{SUB+ENV}(0)=\rho_{SUB} (0) \otimes \rho_{ENV}(0)$.
However, when the interaction is tiny (e.g., as in the case of a
diluted environment) the dynamics of $\rho_{SUB}$ becomes
approximately
free from
memory effects \cite{S}
(Markovian), and it is dictated by a FGKLS master
equation \cite{franke} - \cite{lindblad-2} of the form
\begin{equation}
\dot{\rho}=-i \left( H_{eff} \; \rho(t) - \rho(t) H_{eff}^\dagger \right)
+ L[\rho],
\label{lindblad}
\end{equation}
where the so-called Lindbladian,
\begin{equation}
L[\rho]=
\sum_j \left( A^j \rho (A^j)^\dagger - \frac{1}{2} \rho (A^j)^\dagger A^j
- \frac{1}{2} (A^j)^\dagger A^j \rho \right),
\label{lindblad1}
\end{equation}
with the effective Hamiltonian $H_{eff}$ being in general non-Hermitian. From now on, $\rho$ stands for $\rho_{SUB}$. In what follows we restrict
ourselves by consideration of Hermitian Hamiltonians only that results in the conservation of the total probability $\Tr{\rho}$.

The new piece with vanishing trace is linear in $\rho$, and the linear superposition
principle is
still valid for such a quantum system. But it is
{\it quadratic}\footnote{The form (\ref{lindblad1}) is the most general
Hermitian expression bilinear in $A^j$ with vanishing $Tr (L(\rho)).$ The forms linear in operators $A^j$ are not interesting since they can be absorbed into
Schr\"odinger-Liouville dynamics (\ref{lindblad}) with $L=0.$} in the unspecified operators $A^j$: the FGKLS equation
encodes a new dynamics that is not purely Hamiltonian. The new term may
entail decoherence by making pure states to evolve into mixed states.

If the opposite process does not occur, i.e., no mixed state
evolves to a pure one, the von Neumann entropy
$${\bf S} = \Tr{-\rho \log \rho}$$
should not decrease and it imposes the constraints on the choice of
the FGKLS operators $A^j$. While this may be useful to model the influence of the measurement apparatus, in general case the entropy may decrease and the evolution of the mixed state may end with a pure state \cite{diehl}. In this paper we do not put any restriction on the FGKLS operators.

\section{Supersymmetrical FGKLS system}
The technique of Supersymmetrical Quantum Mechanics and the related isospectral Darboux
transformations \cite{reviews} - \cite{reviews-5}, \cite{ABI}, \cite{ABI-2} are well designed to the $\rho$ matrix formalism.

Indeed, in the fermion number representation, the
one-dimensional  SUSY QM  assembles a pair of
isospectral Hamiltonians
$h_1$ and $h_2$ into
a Super-Hamiltonian,
\ba
H = \left(\begin{array}{cc}
h_1& 0\\
0 & h_2
\end{array}\right) =\left(\begin{array}{cc}
-\partial^2 + V_1(x)& 0\\
0 & - \partial^2 + V_2(x)
\end{array}\right) \equiv -\partial^2 {\bf I} + {\bf V}(x) , \la{suham}
\ea
where $\partial \equiv d/dx$ and the potentials are taken real.
We assume also the {\it time independence} of the
Super-Hamiltonian and also of supercharges which are introduced below.
The isospectral connection between
components of the Super-Hamiltonian is provided by  intertwining
relations with the help of Crum-Darboux
differential  operators\footnote{These differential operators may be of different order in derivatives. Both partner potentials $V_{1,2}(x)$ can be expressed explicitly in terms of coefficients ("superpotentials") of intertwining operators $q^{\pm}$ but the direct expression of $V_2$ through $V_1$ is as a rule impossible \cite{reviews-4}.}  $q^+ = (q^-)^\dagger$ ,
\be
h_1 q^+ = q^+ h_2 , \quad q^- h_1 = h_2 q^- , \la{intertw}
\ee
which, in the framework of SUSY QM,  are components of the supercharges,
\ba
Q=\left(\begin{array}{cc}
 0 &  q^+\\
0 & 0
\end{array}\right),\quad
\bar Q=\left(\begin{array}{cc}
 0 &  0\\
q^- & 0
\end{array}\right) = Q^\dagger,\quad  Q^2 = \bar Q^2 = 0. \la{such}
\ea
The isospectral transformation \gl{intertw} entails the
conservation of supercharges or the supersymmetry of the Super-Hamiltonian,
\be
[H,Q] = [H,\bar Q] =0, \la{cons}
\ee
which represents the basis of the SUSY algebra. Its algebraic closure
is given, in general, by  a non-linear SUSY relation \cite{ais},
\be
\left\{Q,\bar Q\right\} = {\cal P} (H)=
\left(\begin{array}{cc}
 {\cal P} (h_1) &  0\\
0 & {\cal P} (h_2)
\end{array}\right), \la{poly}
\ee
where ${\cal P} (H)$ is a time-independent function of the Super-Hamiltonian .

Let us consider first the Schr\"odinger-Liouville dynamics (\ref{lindblad}), i.e. with vanishing Lindbladian $L[\rho]\equiv 0,$ for the system with the Hamiltonian $h_1$ and
corresponding density matrix $\rho_1,$
\be
\dot{\rho_1}=-i \left( h_1 \ \rho_1(t) - \rho_1(t) h_1 \right). \la{rho-1}
\ee
After multiplying Eq.(\ref{rho-1})
by $q^-$ from the left side
and by $q^+$ from the right one, in the absence of decoherence effects
one obtains,
\be
\dot{\rho_2}=-i \left( h_2 \ \rho_2(t) - \rho_2(t) h_2 \right), \la{rho-}
\ee
where by definition,
\be
\sqrt{{\cal P} (h_2)}\rho_2 \sqrt{{\cal P} (h_2)} = q^- \rho_1 q^+. \la{trans}
\ee
After multiplying of \gl{trans} by $q^+$ from the left or  by $q^-$
from the right and using intertwining relations (\ref{intertw}) one arrives to the equivalent set of intertwining
relations for $\rho_1$ and $\rho_2$,
\ba
&&\sqrt{{\cal P} (h_1)}\rho_1 q^+ = q^+ \rho_2 \sqrt{{\cal P} (h_2)};\nonumber\\
&&\sqrt{{\cal P} (h_2)}\rho_2 q^- = q^- \rho_1 \sqrt{{\cal P} (h_1)}. \la{rhoint}
\ea
Thus the supersymmetrization is perfectly compatible with the $\rho$
matrix approach but it is realized by essentially nonlocal
transformations \gl{trans}. The inverse mapping also holds,
\be
\sqrt{{\cal P} (h_1)}\rho_1 \sqrt{{\cal P} (h_1)} = q^+ \rho_2 q^-\ , \la{transinv}
\ee
which is provided also by Eqs. \gl{rhoint}. Due to intertwining relations (\ref{intertw}) and factorization (\ref{poly}), one can derive from (\ref{trans}) and (\ref{transinv}):
\ba
\rho_1&=& \sum^{M_1}_{j,k=1}\beta_{1}^{jk} |\psi_-^j\rangle\langle\psi_-^k| + q^+\frac{1}{\sqrt{{\cal P} (h_2)}}\rho_2 \frac{1}{\sqrt{{\cal P} (h_2)}}q^-,   \label{rrho}\\
\rho_2&=& \sum^{M_2}_{j,k=1}\beta_{2}^{jk} |\psi_+^j\rangle\langle\psi_+^k| + q^-\frac{1}{\sqrt{{\cal P} (h_1)}}\rho_1 \frac{1}{\sqrt{{\cal P} (h_1)}}q^+,   \label{rrhoo}
\ea
where $|\psi_{\pm}^i\rangle$ are normalized zero modes of the operators $q^{\pm},$ respectively, with the matrices $\beta_{1,2}^{jk}= \langle\psi_{\mp}^j|\rho_{1,2}|\psi_{\mp}^k\rangle$. Here and for the rest of the paper we understand the inverse operators of ${\cal P}(h_{1,2})$ in a sense that they are supplemented by projectors on the subspaces orthogonal to the zero modes i.e.
\be
\frac{1}{\sqrt{{\cal P} (h_1)}}|\psi_-^j\rangle\equiv 0,\quad
\frac{1}{\sqrt{{\cal P} (h_2)}}|\psi_+^j\rangle\equiv 0,
\label{invsquare}
\ee
Note that the relations \gl{rrho},\gl{rrhoo} prohibit the mixing between the zero modes and positive energy states in the matrices $\rho_1$ and $\rho_2$. Another consequence is that,
\be
\Tr{\rho_1}-\Tr{\beta_1}=\Tr{\rho_2}-\Tr{\beta_2},\la{str}
\ee
thus the total probability is invariant if $\Tr{\beta_1}=\Tr{\beta_2}$.

The Super-density matrix $\hat{\rho}$ defined as
\be
\hat\rho \equiv\left(\begin{array}{cc}
 \rho_1 &  0\\
0 & \rho_2
\end{array}\right) \label{hatrho}
\ee
satisfies the following linear equation:
\be
\sqrt{{\cal P} (H)}\hat\rho \sqrt{{\cal P} (H)}
=  Q \hat\rho \bar Q + \bar Q \hat\rho Q  , \la{superho1}
\ee
which is an essence of the SUSY for $\rho$ matrices.
Equivalently the set \gl{rhoint} is reproduced by,
\be
 \bar Q \hat\rho\sqrt{{\cal P} (H)}
= \sqrt{{\cal P} (H)}  \hat\rho \bar Q, \quad
Q \hat\rho\sqrt{{\cal P} (H)}
= \sqrt{{\cal P} (H)}  \hat\rho Q, \label{15}
\ee
which are analogs of the intertwining relations (\ref{intertw}).

At this point,
one may find it useful to introduce normalized supercharges,
\be
 Q_n \equiv Q \frac{1}{\sqrt{{\cal P} (H)}}
= \frac{1}{\sqrt{{\cal P} (H)}} Q;
\quad \bar Q_n \equiv \bar Q \frac{1}{\sqrt{{\cal P} (H)}}
= \frac{1}{\sqrt{{\cal P} (H)}} \bar Q, \la{normch}
\ee
which satisfy the following superalgebra:
\be
 \{Q_n, \bar Q_n\} = 1;\quad  [Q_n, \hat\rho] = [\bar Q_n, \hat\rho] = 0; \quad \hat\rho =  Q_n \hat\rho \bar Q_n + \bar Q_n \hat\rho Q_n, \la{superho2}
\ee
the latter relation being actually a direct consequence of the former one (\ref{superho1}).
Thus, the $\hat\rho$ matrix is supersymmetric in terms of normalized supercharges.

Now let us derive the SUSY covariance conditions on the FGKLS operators $A^j_1\,\leftrightarrow \,A^j_2$
associated with the Hamiltonians $h_1,\,h_2$.
It is clear that the definition similar to Eq.(\ref{trans}):
\be
q^- \, L[A^j_1; \rho_1]\ q^+ =  \sqrt{{\cal P} (h_2)}\,
L[A^j_2; \rho_2]\, \sqrt{{\cal P} (h_2)}.
\ee
will provide the analogous SUSY covariance properties for all terms in the FGKLS master equation (\ref{lindblad}).
It entails the following (similar to (\ref{rhoint})) intertwining relations between the FGKLS operators,
\ba
&& q^- A^j_1 \sqrt{{\cal P} (h_1)} =
\sqrt{{\cal P} (h_2)} A^j_2 q^-;\label{20}\\
&& q^- \sum_j (A^j_1)^\dagger A^j_1 \sqrt{{\cal P} (h_1)} =
\sqrt{{\cal P} (h_2)}\sum_j (A^j_2)^\dagger A^j_2 q^-. \label{2020}
\ea
Multiplying (\ref{20}) on the left by $q^+$  and on the right by $ 1/\sqrt{{\cal P} (h_2)}$ and using intertwining relations (\ref{intertw}), one obtains
explicit expressions:
\be
A^j_2=q^-\frac{1}{\sqrt{{\cal P} (h_1)}}A^j_1\frac{1}{\sqrt{{\cal P} (h_1)}}q^+;\quad
A^j_1=q^+\frac{1}{\sqrt{{\cal P} (h_2)}}A^j_2\frac{1}{\sqrt{{\cal P} (h_2)}}q^-.
\label{A-minus}
\ee
Simultaneously, the relations (\ref{2020}) also will be fulfilled.

When zero modes of $q^\pm$ are present \gl{A-minus} is no longer an unique solution of \gl{20} and \gl{2020}. However the corresponding evolution does not introduce mixing between the zero modes and positive energy sector in $\rho_1$ and $\rho_2$ that would contradict \gl{rrho} and \gl{rrhoo}. Thus in the rest of the paper we restrict ourselves to this special solution.

One can also introduce the SUSY notations,
\ba
&& \hat A^j \equiv\left(\begin{array}{cc}
 A^j_1 &  0\\
0 & A^j_2
\end{array}\right),\label{22}\\
&&  \bar Q \hat A^j \sqrt{{\cal P} (H)}
= \sqrt{{\cal P} (H)} \hat A^j \bar Q, \quad
\bar Q \sum_j (\hat A^j)^{\dagger} \hat A^j \sqrt{{\cal P}(H)}
= \sqrt{{\cal P} (H)} \sum_j (\hat A^j)^{\dagger} \hat A^j \bar Q \label{2222}\\
&& Q (\hat A^j)^{\dagger} \sqrt{{\cal P} (H)}
= \sqrt{{\cal P} (H)} (\hat A^j)^{\dagger} Q,\quad
Q \sum_j (\hat A^j)^{\dagger} \hat A^j \sqrt{{\cal P} (H)}
= \sqrt{{\cal P} (H)} \sum_j (\hat A^j)^{\dagger} \hat A^j Q. \la{intlin1}
\ea
If one employs the normalized SUSY charges \gl{normch} then the above
relations appear to be intertwining ones,
\ba
&&  \bar Q_n \hat A^j
= \hat A^j \bar Q_n, \quad
\bar Q_n \sum_j (\hat A^j)^{\dagger} \hat A^j
=  \sum_j (\hat A^j)^{\dagger} \hat A^j \bar Q_n \label{23}\\
&& Q_n (\hat A^j)^{\dagger}
= (\hat A^j)^{\dagger} Q_n,\quad
Q_n \sum_j (\hat A^j)^{\dagger} \hat A^j
=  \sum_j (\hat A^j)^{\dagger} \hat A^j Q_n. \la{intlin2}
\ea
For hermitian $\hat A^j$ the relations \gl{intlin1}, \gl{intlin2}
are reduced to the following equations,
\ba
&&\sqrt{{\cal P} (H)} \hat A_j \sqrt{{\cal P} (H)}
= Q \hat A_j \bar Q +  \bar Q \hat A_j Q,\quad
 \hat A_j
= Q_n \hat A_j \bar Q_n +  \bar Q_n \hat A_j Q_n,\nonumber \\
&&\sqrt{{\cal P} (H)} \sum_j (\hat A_j)^2 \sqrt{{\cal P} (H)}
= Q  \sum_j (\hat A_j)^2 \bar Q +
\bar Q \sum_j (\hat A_j)^2 Q,\nonumber \\
&&\sum_j (\hat A_j)^2
= Q_n  \sum_j (\hat A_j)^2\bar Q_n +  \bar Q_n \sum_j (\hat A_j)^2 Q_n,\label{39}
\ea
which are simple consequences of the intertwining conditions (\ref{2222}),
\gl{intlin1}, \gl{intlin2}. We remark that the latter relations
remind the similar ones
\gl{superho1},\gl{superho2} for the Super-density matrix $\hat\rho.$ Thus both the $\hat\rho$-matrix and
the (hermitian) matrices $\hat A_j$ are solutions of the same set of equations.
On the other hand, the Super-density matrix $\hat\rho$ is a positive operator whereas there is no reason
for the FGKLS operators to be positive.

\section{Examples}
In this Section, we shall consider a few simple illustrative examples of SUSY construction proposed above. Let us consider the density matrix
$\rho_1$ for one-dimensional quantum system in the energetic basis:
\be
h_1|E_m\rangle = E_m|E_m\rangle, \label{0}
\ee
which for simplicity has only discrete spectrum states with non-degenerate mutually orthonormal  eigenvectors $|E_m\rangle$. The density matrix $\rho_1$ \gl{rrho} can be represented in the energy basis as,
\be
\rho_1(t) = \sum_{m,n=0}^{\infty} r_{mn} |E_m\rangle \langle E_n|+\beta_1|0\rangle\langle0|.  \label{1+}
\ee
where $|0\rangle$ is a possible zero mode of $q^-$ and $r_{mn}$ is a Hermitian semipositively definite matrix. Because the solution \gl{A-minus} acts trivially on the zero mode sector we will omit it from our consideration by assuming $\beta_1=0$. Also, let the components $q^{\pm}$ of supercharges be the differential operators of first order so that the polynomials ${\cal P} (H)$ in (\ref{poly}) and further on are linear in the corresponding Hamiltonians with zero energy of factorization: ${\cal P} (h_1)=q^+q^-=h_1 ,\quad {\cal P} (h_2)=q^-q^+=h_2.$
Under the supertransformations, the eigenstates $|E_m\rangle$ of $h_1$ are transformed to the eigenstates $|\widetilde E_m\rangle$ of $h_2:$
\be
|\widetilde E_m\rangle =\frac{1}{\sqrt{E_m}} q^- |E_m\rangle; \quad h_2 |\widetilde E_m\rangle = \widetilde E_m |\widetilde E_m\rangle. \label{6}
\ee
Excluding the possible zero modes the spectra of $h_1$ and $h_2$ coincide exactly $\widetilde E_n = E_n$ (see details in \cite{ACDI-1} - \cite{ACDI-1-3}). It follows from (\ref{rrhoo}) and (\ref{6}) that
\be
\rho_2(t) = \sum_{m,n=0}^{\infty} r_{mn} |\widetilde E_m\rangle \langle\widetilde E_n|.  \label{7}
\ee

1. Let us choose the FGKLS operators $A_1^j$ as the real functions of the Hamiltonian:
\be
A_1^j=f_j(h_1);\quad f_j=f_j^{\star}. \label{3}
\ee
The FGKLS equation (\ref{lindblad}), (\ref{lindblad1}) for the open system with $h_1,\, A^j_1$ takes the form:
\ba
&&\sum_{n,m=0}^{\infty}\frac{d}{dt}r_{mn} |E_m\rangle\langle E_n| = \label{5}\\
&&= \sum_{n,m=0}^{\infty} r_{mn} \biggl[ -i(E_m-E_n) -
\frac{1}{2}\sum_j (f_j(E_n)-f_j(E_m))^2 \biggr] |E_m\rangle\langle E_n|,  \nonumber
\ea
providing the necessary relations for $r_{mn}(t).$ In particular, the asymptotically stable at $t\to\infty$ density matrix is possible if for $n \neq m$
at $t\rightarrow\infty $ the matrix $r_{mn} \rightarrow 0.$ The diagonal elements $r_{nn}$ may be arbitrary up to
normalization condition (\ref{2}) due to vanishing expression of terms with $n=m$ in square brackets in (\ref{5}).
Supersymmetry between the partners $\rho_1, \, \rho_2 $ leads to conclusion that the partner density matrix $\rho_2(t)$ is also asymptotically stable if its FGKLS operators are taken according to supersymmetry as in Eq.(\ref{A-minus}): $A_2^j = f_j(h_2).$

2. One more example can be considered with
\be
A_1(t) = \sum_{s=1}^{2}\gamma_s|E_s\rangle\langle E_s|
; \quad A_1^{\dagger}(t) = \sum_{s=1}^{2}\gamma_s^{\star}|E_s\rangle\langle E_s|
, \label{11}
\ee
with complex values of $\gamma_{1,2}(t).$ According to (\ref{A-minus}), the superpartners are:
\be
A_2(t) =\sum_{s=1}^{2} \gamma_s|\widetilde E_s\rangle\langle\widetilde E_s|
; \quad A_2^{\dagger}(t) = \sum_{s=1}^{2} \gamma_s^{\star}|\widetilde E_s\rangle\langle\widetilde E_s|
, \label{12}
\ee
with the same functions $\gamma_{1,2}(t)$ and $|\widetilde E\rangle$ defined by (\ref{6}). Such choice for the FGKLS operators leads to the Lindbladian:
\ba
L[\rho_1, A_1] &=& \sum_{s,u=1}^{2} \gamma_s r_{su}\gamma^{\star}_{u}|E_s\rangle\langle E_u| -
\frac{1}{2}\sum_{s,u=1}^{2}\sum_{n=0}^{\infty} r_{ns}|\gamma_s|^2|E_n\rangle\langle E_s| -  \nonumber\\
&-&\frac{1}{2}\sum_{s,u=1}^{2}\sum_{m=0}^{\infty} r_{sm}|\gamma_s|^2|E_s\rangle\langle E_m|  .  \label{13}
\ea
Since both the eigenenergies $\widetilde E_n=E_n$ and functions $\gamma_s(t)$ for the FGKLS equations for superpartners $\rho_1(t),\,\rho_2(t)$ are the same,
the evolution of density matrices is also the same while the Hamiltonians $h_1,\,h_2$ are different. This fact opens an opportunity to predict evolution of $\rho_2$ if evolution of $\rho_1$ is already known.

3. In the next example, we assume that neither $q^-$ nor $q^+$ has zero modes and use the only FGKLS operator $A_1=q^- \frac{\Delta}{\sqrt{h_1}}$ where the real constant $\Delta$ is of dimensionality $[\Delta]=[E^{1/2}].$ Then, by relation (\ref{A-minus}):
\be
A_1=A_2=q^-\frac{\Delta}{\sqrt{h_1}}=\frac{\Delta}{\sqrt{h_2}}q^-; \quad A_1^{\dagger}=A_2^{\dagger}=\frac{\Delta}{\sqrt{h_1}}q^+=q^+\frac{\Delta}{\sqrt{h_2}}. \label{9}
\ee
for which the Lindbladians in the energetic basis are:
\ba
L[\rho_1, A_1]&=&\Delta^2\sum_{m,n=0}^{\infty} r_{mn}
\biggl[ |\widetilde E_m\rangle \langle\widetilde E_n| - |E_m\rangle \langle E_n| \biggr] = \Delta^2 (\rho_2 - \rho_1);    \label{10}\\
L[\rho_2, A_2]&=& \Delta^2\sum_{m,n=0}^{\infty} r_{mn}
\biggl[ \frac{1}{\sqrt{h_2}} q^- |\widetilde E_m\rangle \langle\widetilde E_n|q^+\frac{1}{\sqrt{h_2}} - |\widetilde E_m\rangle \langle\widetilde E_n| \biggr].
\label{10+}
\ea
Taking into account, that $\widetilde E_n = E_n$ here, the FGKLS equations for superpartners $\rho_1,\,\rho_2$ can be written now as:
\ba
\dot\rho_1(t)
&=& \sum_{m,n=0}^{\infty} r_{mn}
\biggl[ (-i)(E_m-E_n-i\Delta^2)| E_m\rangle \langle E_n| + \Delta^2 |\widetilde E_m\rangle \langle\widetilde E_n| \biggr]; \label{10-1}\\
\dot\rho_2(t)& =& \sum_{m,n=0}^{\infty} r_{mn}
\biggl[ (-i)(E_m - E_n-i\Delta^2)|\widetilde E_m\rangle \langle\widetilde E_n| +  \frac{\Delta}{\sqrt{h_2}} q^- |\widetilde E_m\rangle \langle\widetilde E_n|q^+\frac{\Delta}{\sqrt{h_2}}\biggr]
\label{10-2}
\ea

Using the following representation,
\be
\frac{1}{\sqrt{h_2}}q^-=\sum_{n=0}^{\infty}|\tilde{E}_n\rangle\langle E_n|,\quad
\frac{1}{\sqrt{h_2}}q^+=\sum_{n=0}^{\infty}|E_n\rangle\langle \tilde{E}_n|.
\label{qdecomp}
\ee
one can demonstrate that both equations (\ref{10-1}) and (\ref{10-2}) lead to the same evolution of the coefficients $r_{mn}$,
\be
\frac{d}{dt}r_{mn}=-ir_{mn}(E_m-E_n-i\Delta^2)
+\Delta^2\sum_{k,l=0}^{\infty}r_{kl}
\langle E_m|\tilde{E}_k\rangle
\langle \tilde{E}_l|E_n\rangle.
\label{rhoevo}
\ee

4. The previous example is easily generalized on the case when $q^-$ has a zero mode $|0\rangle$. Then the FGKLS operator $A_1$ should be supplemented with the projector on the orthogonal subspace,
\be
A_1=\Big(1-|0\rangle\langle 0|\Big)q^-\frac{\Delta}{\sqrt{h_1}},\quad
A_1^\dagger=\frac{\Delta}{\sqrt{h_1}}q^+\Big(1-|0\rangle\langle 0|\Big),
\ee
whereas its superpartner remains to be the same,
\be
A_2=q^-\frac{\Delta}{\sqrt{h_1}},\quad
A_2^\dagger=\frac{\Delta}{\sqrt{h_1}}q^+.
\ee

This modification happens to be non-trivial if at least some $\langle \tilde{E}_n|0\rangle\neq 0$. However the Lindbladian $L[\rho_2,A_2]$ preserves the form (\ref{10+}). As result the evolution of the coefficients $r_{mn}$ is given by (\ref{rhoevo}).

This specific example includes the case of harmonic oscillator when components of supercharges coincide with creation-annihilation operators $q^-=a,\, q^+=a^\dagger$ playing also the role of FGKLS operators.

\section{Conclusions}
In the present paper, we investigated FGKLS equation for a pair of SUSY-partner Hamiltonian systems. This allowed us to generalize the SUSY QM transformations initially written for the stationary Schr\"odinger equation to the case of open quantum systems described by FGKLS equations with corresponding Lindblad operators. In the last Section, several illustrative examples of this algorithm demonstrated the SUSY-partnership of different open systems which both satisfy master FGKLS equation. It is clear that similarly to the chain of successive SUSY transformations of the Hamiltonians in standard SUSY QM \cite{reviews-4}, \cite{ais} - \cite{levels-2}, the iteration of SUSY transformations (\ref{trans}) of the density matrices can be performed both in the general form and for examples of the previous Section.

While we restricted ourselves to the stationary case the supersymmetric approach may be applied to the nonstationary systems. For example, it is useful to mention the extension of standard SUSY QM to nonstationary Schr\"odinger equation and to classical stochastic dynamical systems characterized by a Fokker-Planck equation (see \cite{nature-8}, \cite{junker}).

In the recent paper \cite{ilievsky} the structure of the steady-state ("pointer") solutions of open integrable quantum lattice models with boundaries, driven far from equilibrium by incoherent particle reservoirs attached at the boundaries, was investigated. Just this class of models was generalized to the graded $SU(n|m)$ chains, which represent interacting integrable models with fermionic degrees of freedom. In the present paper, we study SUSY transformations for arbitrary (not only steady-state solutions) open systems without any restrictions on integrability or exact solvability of an initial Hamiltonian. Thus, we obtain the convenient tool for the quantum engineering of open systems described by the FGKLS equation: starting from the simple enough open system, one can obtain the partner system (or, even the chain of systems) with known properties.
Also, our approach opens perspectives to generate/eliminate dark states in density matrices which play a crucial role in governing the behavior of entropy.

\section{Acknowledgments}
The work was supported by RFBR Grant No. 18-02-00264-a. The authors are grateful to all of multiple anonymous referees for useful comments.


\begin{thebibliography}{99}
\bibitem{franke}
Franke V.A.
1976 {\it Theor. Math. Phys.} {\bf 27} 406
\bibitem{sudar}
Gorini V., Kossakowski A., and Sudarshan E.C.G.
1976 {\it J. Math. Phys.} {\bf 17} 821
\bibitem{lindblad}
Lindblad G.
1976 {\it Comm. Math. Phys.} {\bf 48} 119
\bibitem{lindblad-2}
Lindblad G.
1976 {\it Rep. Math. Phys.} {\bf 10} 393
\bibitem{history}
Chruscinski D. and Pascazio S.
2017 {\it Open Systems and Information Dynamics} {\bf 24} 1740001 (arXiv:1710.05993)
\bibitem{franke-2}
Kurkov M.A. and V.A.Franke V.A.
2011 {\it Found. Phys.} {\bf 41} 820
\bibitem{PH}
Huet P. and Peskin M.
1995 {\it Nucl. Phys.} {\bf B434} 3
\bibitem{BF}
Benatti F. and Floreanini R.
1997 {\it Nucl. Phys.} {\bf B488} 335
\bibitem{E}
Ellis J., Mavromatos N.E. and Nanopoulos D.V.
1992 {\it Phys. Lett.} {\bf B293} 142
\bibitem{E-2}
Ellis J., Lopez J.L., Mavromatos N.E. and Nanopoulos D.V.
1996 {\it Phys. Rev.} {\bf D53} 3846
\bibitem{ATT}
Andrianov A.A., Taron J. and Tarrach R.
2001 {\it Phys.Lett.} {\bf B507} 200
\bibitem{many}
Breuer F. and Petruccione H.-P.
2002 {\it Theory of Open Quantum Systems} (Oxford University Press, New York)
\bibitem{many-2}
2003 Irreversible Quantum Dynamics (Lecture Notes in Physics vol 622) ed. F.Benatti and R.Floreanini (Springer, Berlin, Heidelberg)
\bibitem{many-3}
Tarasov V.E.
2008 {\it Quantum Mechanics of Non-Hamiltonian and Dissipative Systems} (Elsevier, Amsterdam)
\bibitem{albert}
Albert V.V. and Liang Jiang
2014 {\it Phys. Rev.} {\bf A89} 022118
\bibitem{albert-2}
Tomka M., Pletyukhov M. and Gritsev V.
2015 {\it Sci. Rep.} {\bf 5} 13097
\bibitem{albert-3}
Macho A., Llorente R. and Garcia-Meca C.
2018 {\it Phys. Rev. Appl.} {\bf 9} 014024
\bibitem{weinberg}
Weinberg S.
2012 {\it Phys. Rev.} {\bf A85} 062116
\bibitem{weinberg-2}
Weinberg S.
2014 {\it Phys. Rev.} {\bf A90} 042102
\bibitem{reviews}
Cooper F., Khare A. and Sukhatme U.
1995 {\it Phys. Rep.} {\bf 251} 267
\bibitem{reviews-2}
Bagchi B.K.
2001 {\it Supersymmetry in Quantum and Classical Mechanics} (Chapman, Boca Raton)
\bibitem{reviews-3}
Fernandez C D.J.
2010 Supersymmetric Quantum Mechanics {\it AIP Conf. Proc.} {\bf 1287} 3
\bibitem{reviews-4}
Andrianov A.A. and Ioffe M.V.
2012 {\it J. Phys.} {\bf A45} 503001
\bibitem{reviews-5}
Fernandez C D.J.
Fernandez C D J 2018 {\it Integrability, Supersymmetry and Coherent States. A Volume in Honour of Professor Veronique 
Hussin (CRM Series in Mathematical Physics)} ed S Kuru et al pp 37–68;
(arXiv:1811.06449)
\bibitem{nature}
Andrianov A.A., Borisov N.V. and Ioffe M.V.
1984 {\it Sov. Phys. JETP Lett.} {\bf 39} 93
\bibitem{nature-2}
Kostelecky V.A. and Nieto M.M.
1984 {\it Phys. Rev. Lett.} {\bf 53} 2285
\bibitem{nature-3}
Comtet D.K.C.A. and Bandrauk A.
1985 {\it Phys. Lett.} {\bf B150} 159
\bibitem{nature-4}
Andrianov A.A. and Ioffe M.V.
1991 {\it Phys. Lett.} {\bf B255} 543
\bibitem{nature-5}
Gozzi E., Reuter M. and Thacker W.D.
1993 {\it Phys. Lett.} {\bf A183} 29
\bibitem{nature-6}
Andrianov A.A., Cannata F., Ioffe M.V. and Nishnianidze D.N.
1997 {\it J. Phys.} {\bf A30} 5037
\bibitem{nature-7}
Andrianov A.A., Cannata F., Dedonder J.-P. and Ioffe M.V.
1999 {\it Int. J. Mod. Phys.} {\bf A14} 2675
\bibitem{nature-8}
Cannata F., Ioffe M., Junker G. and Nishnianidze D.,
1999 {\it J. Phys.} {\bf A32} 3583
\bibitem{nature-9}
Cannata F., Ioffe M.V., Neelov A.I. and Nishnianidze D.N.
2004 {\it J. Phys.} {\bf A37} 10339
\bibitem{nature-10}
Ioffe M.V., Kuru S., Negro J. and Nieto L.M.
2006 {\it J. Phys.} {\bf A39} 6987
\bibitem{ais}
Andrianov A.A., Ioffe M.V. and Spiridonov V.
1993 {\it Phys. Lett.} {\bf A174} 273
\bibitem{ACDI-1}
Andrianov A.A., Cannata F., Dedonder J.-P. and Ioffe M.V.
1995 {\it Int. J. Mod. Phys.} {\bf A10} 2683
\bibitem{levels-2}
Andrianov A.A. and Sokolov A.V.
2003 {\it Nucl. Phys.} {\bf B660} 25
\bibitem{levels}
M. V. Ioffe and D. N. Nishnianidze,
Phys. Lett. A327 (2004) 425
\bibitem{S}
Spohn H.
1980 {\it Rev. Mod. Phys.} {\bf 52} 569
\bibitem{diehl}
Diehl S., Micheli A., Kantian A., Kraus B., B{\"u}chler H.P., and Zoller P.
2008 {\it Nature Physics} {\bf 4} 878
\bibitem{ABI}
Andrianov A.A., Borisov N.V. and Ioffe M.V.
1984 {\it Phys. Lett.} {\bf A105} 19
\bibitem{ABI-2}
Andrianov A.A., Borisov N.V. and Ioffe M.V.
1984 {\it Theor. Math. Phys.} {\bf 61} 1078
\bibitem{ACDI-1-2}
Samsonov B.F.
1999 {\it Phys. Lett.} {\bf A263} 274
\bibitem{ACDI-1-3}
Aoyama A., Sato M. and Tanaka T.
2001 {\it Phys. Lett.} {\bf B503} 423
\bibitem{junker}
Junker G.
1996 {\it Supersymmetric methods in Quantum and Statistical Physics} (Berlin: Springer) (Chapter 7)
\bibitem{ilievsky}
Ilievski E.
2017 {\it SciPost. Phys.} {\bf 3} 031

\end{thebibliography}
\end{document}